\documentclass[acmlarge]{acmart}

\makeatletter
\newcommand{\myconfshort}{\acmConference@shortname}
\newcommand{\myconffull}{\acmConference@name}
\newcommand{\myconfdate}{\acmConference@date}
\newcommand{\myconfloc}{\acmConference@venue}
\AtBeginDocument{
  \fancypagestyle{firstpagestyle}{
    \fancyhead{}%
    \fancyfoot[C]{}%
  }
  \fancyhf{}
  \fancyhead[LO]{\@headfootfont\shorttitle}%
  \fancyhead[RE]{\@headfootfont\@shortauthors}%
  \fancyhead[LE]{\@headfootfont\footnotesize \myconfshort, \myconfdate, \myconfloc}%
  \fancyhead[RO]{\@headfootfont\footnotesize \myconfshort, \myconfdate, \myconfloc}%
  \fancyfoot[C]{}%
}
\makeatother
\acmBooktitle{\conffull\@ (\confshort), \confdate, \confloc}
\AtBeginDocument{%
  }

\usepackage{hyperref}

\usepackage{tcolorbox}

\setcopyright{acmlicensed}

\copyrightyear{2026}
\acmYear{2026}
\setcopyright{cc}
\setcctype{by}
\acmConference[FAccT '26]{The 2026 ACM Conference on Fairness, Accountability, and Transparency}{June 25--28, 2026}{Montreal, QC, Canada}
\acmBooktitle{The 2026 ACM Conference on Fairness, Accountability, and Transparency (FAccT '26), June 25--28, 2026, Montreal, QC, Canada}
\acmDOI{10.1145/3805689.3812228}
\acmISBN{979-8-4007-2596-8/2026/06}

\begin{document}

\title[Auditing African Content Moderators' Working Conditions]{Auditing African Content Moderators' Working Conditions by Using the European General Data Protection Regulation (GDPR)}


\author{Mariame Tighanimine}
\authornote{Equal contribution}
\email{mariame@tighanimine.com}
\orcid{0009-0005-8188-9976}
% \authornotemark[2] 
%\authornote{Also affiliated with Lise, Cnam CNRS, France.}
\authornote{Corresponding author: mariame@tighanimine.com}
\affiliation{%
  \institution{University of Neuchâtel}
  \country{Switzerland}}
\affiliation{%
  \institution{PersonalData.IO}
  \country{Switzerland}}
\affiliation{%
  \institution{Lise Cnam CNRS}
  \country{France}}
\author{Jessica Pidoux}
\email{jessica.pidoux@unine.ch}
\orcid{0000-0001-5705-6230}
\authornotemark[1]
\affiliation{%
  \institution{University of Neuchâtel}
  \country{Switzerland}}
\affiliation{%
  \institution{PersonalData.IO}
  \country{Switzerland}}

\author{Sonia Kgomo}
\email{soniamatete35@gmail.com}
\orcid{0009-0008-7865-6962}
\affiliation{%
  \institution{African Content Moderators Union}
  \country{Kenya}}

\author{Kauna Ibrahim Malgwi}
\email{kaunamalgwi@gmail.com}
\orcid{0009-0008-9685-740X}
\affiliation{%
  \institution{African Content Moderators Union}
  \country{Kenya}}
\affiliation{%
  \institution{Kenya Digital Rights and Mental Health Initiative}
  \country{Kenya}}

\author{Richard Mwaura Mathenge}
\email{richiemathenge9@gmail.com}
\orcid{0009-0004-7100-0259}
\affiliation{%
  \institution{African Content Moderators Union}
  \country{Kenya}}
\affiliation{%
  \institution{Kenya Techworker Community Africa}
  \country{Kenya}}

\author{Mophat Okinyi}
\email{okinyi.mophat@gmail.com}
\orcid{0009-0000-5932-3232}
\affiliation{%
  \institution{African Content Moderators Union}
  \country{Kenya}}
\affiliation{%
  \institution{Kenya Techworker Community Africa}
  \country{Kenya}}

\author{James Oyange}
\email{oyangej8@gmail.com}
\orcid{0009-0009-4917-710X}
\affiliation{%
  \institution{African Content Moderators Union}
  \country{Kenya}}

\renewcommand{\shortauthors}{Tighanimine et al.}

\begin{abstract}

In this article, we audit the working conditions of content moderators in Kenya and Nigeria employed by business process outsourcing (BPO) companies by using the European General Data Protection Regulation (GDPR). We demonstrate its extraterritorial scope for gaining access to elements such as employment contracts and NDAs that have never been provided to the workers concerned. The results of this approach provide legally grounded evidence of the structural disadvantages faced by content moderators in the Global South, whose exploitative working conditions violate workers' rights. Our work also highlights the benefits of legislation aimed at protecting individuals' data rights as a counterweight to the tech industry's discourse of exceptionalism, which obscures its dependence on BPOs to externalise labour costs and accountability, whilst claiming that its products, business models, and methods of resource extraction are unprecedented and fall outside any existing legal framework.

\end{abstract}




\maketitle

\section{Introduction}

The systematic concealment of industrial labour within supply chains by companies has persisted throughout history, obscuring workers from public view and enabling the precarisation of their conditions, while simultaneously giving rise to collective demands and social progress. The precarious situation of content moderators in the contemporary AI industry is not unprecedented; rather, it represents the latest manifestation of this longstanding dynamic. Yet the technology industry also exhibits distinct characteristics that warrant specific attention. In recent decades, a narrative of exceptionalism has become prominent in this sector. This socio-legal imaginary treats platform companies, their products and associated business models as disruptive and exceptional, thereby enabling them to engage in regulatory arbitrage \cite{doorn2020price, vallas2020platforms}. Critically, for workers, this implies that they do not fit any form of existing working models. However, numerous studies have shown the precarity, retrograde and exploitative nature of the forms of work in platforms like Uber \cite{berg2018digital, hauben2020platform, Pidoux_Dehaye_Gursky_2024, pidoux2025gaining, bernarduberization} and more generally with actors in the gig economy over the world \cite{anwar2021between, aloisi2022your, pankaj2024gig, humanrightswatchgig}. Recent gig worker's mobilisations \cite{tassinari2020riders, woodcock2021fight, de2023algorithmic} and the adoption of legislation like the EU's Platform Workers' Directive have proven that platform exceptionalism relates more to a well-honed narrative to justify an economy of servitude \cite{o2022farewell}.

In the AI industry, the exceptionalism narrative has gained renewed vigour. The development of increasingly powerful AI models that are presented as exceptional in human history \cite{bender2025ai}—the race toward AGI \cite{blili-hamelin2025position, El-Mhamdi_Hoang_Tighanimine_2025} represents the strongest version of exceptionalist discourse—has been accompanied by reliance on low-wage labour, particularly in countries in the Global South \cite{anwar2022digital}. In recent years, the demand for data \cite{datahunger}—in a culture of machine learning where data work is seen by computer scientists as ``silly work'' \cite{heikkila2023openai} and working on models is perceived as real work—has intensified, considered as a form of ``voracious appetite'' \cite{bloomberg} to the point where questions about compliance with key privacy principles are being raised. Consequently, this raises the question of whether technological exceptionalism serves to justify a regression in fundamental principles and rights at work as defined by the International Labour Organization (ILO).

The aim of our work is to examine the tangible effects of these developments. By focusing on a collective of content moderators based in Africa who have formed a union to defend their interests as workers, denounce their working conditions, and gain recognition for their professionalism, we have co-led participatory research demonstrating that African content moderators are deprived of conditions that guarantee ``decent work'' \cite{work1999international} or more accurately, ``meaningful work'' \cite{rawls2017theory, rawls2020political, rawls1993law}. These workers are denied knowledge of the content of their work in their employment contracts, which is an essential element of the social bases of self-respect \cite{rawls1993law}. They also lack access to a clear and explicit description of their tasks, which would enable them to decide how to perform their work and have a say in the evolution of work processes and company policies \cite{arneson1987meaningful}. Finally, workers were required to sign particularly restrictive non-disclosure agreements (NDAs), of which they were not provided a copy. The institution of the employment contract is considered a major step forward in the historical transition away from forced labour. It made possible the freedom to participate in the labour market, which Sen \cite{sen2014development} considers both a condition for development and a fundamental freedom.

Our methodological approach to deconstructing technological exceptionalism draws on an analysis of content moderators' working conditions, using the European General Data Protection Regulation (GDPR) as an investigative legal tool. By facilitating workers' exercise of their data subject rights, this approach enables an audit of AI technologies from the perspective of those who perform data work: a crucial yet often overlooked component of the AI supply chain. Our work contributes to the research that has been developing in recent years to address issues of fairness, transparency, and accountability in AI systems in a concrete manner. Indeed, precarious and hidden human labor is an integral part of AI systems \cite{workers2025we}. Thus, ``any serious conversation about the ethics and justice of AI must include the workers who make these systems possible \cite{miceli2025methodological}.'' We strive to be consistent on ethical issues, contributing both to documenting the conditions under which AI systems are produced by human workers and to advancing research that directly engages those workers.

This study makes two contributions. First, it is methodological and legal: it demonstrates that workers in the AI supply chain, located in jurisdictions with weak data protection frameworks, can invoke GDPR rights against employers whose data processing activities involve transfers of personal data to Europe, in order to recover employment documents and audit their working conditions. Second, it is empirical: drawing on four cases, we provide legal evidence on working conditions derived from documents transmitted by the companies themselves, offering a rare insider view of labour practices in the AI supply chain.
\section{Context and Motivation: From Coal Mining to Data Work}

What content moderators experience today, including invisibility, job insecurity, low pay, and lack of unionisation \cite{anwar2022digital, cant2024feeding}, is comparable to what coal miners experienced in the early days of industry. Both industries are fundamentally extractive: coal mining extracted raw material to power industrial capitalism, while the technology industry extracts data processed by workers to power AI systems. This labour is substantial: Muldoon et al. \cite{muldoon2024typology} note that data processing, defined as ``human labour required to support machine learning algorithms through the preparation and evaluation of datasets'', represents as much as 80\% of project time on AI models, forming a critical component of the AI supply chain.

The parallel with coal mining illuminates the transformative potential of labour organising. As Tomassetti \cite{tomassetti2022droit, tomassetti2023energy} describes, miners' unions contributed to constructing the fundamental categories of labour law \cite{webb1920history} and helped shape the institutions characterising contemporary democracies \cite{mitchell2009carbon, deakin2011contribution}, initially mobilising around work related demands \cite{church2002strikes, beynon2024shadow} before extending to broader social and political issues.

The contemporary situation of content moderators in Africa \cite{anwar2020digital, anwar2022digital}, Latin America \cite{miceli2022data, posada2022embedded}, and Asia \cite{chen2022workers, chandhiramowuli2024making, wu2025global}, where work is outsourced to business process outsourcing (henceforth BPO) companies \cite{miceli2022data, cant2024feeding, muldoon2025poverty}, can be understood within this global labour history. However, violations of content moderators' rights have received limited attention. Illiberal political contexts facilitate exploitation \cite{NYTbigtech, techcorruption, Amnestybigtech}, with data work perceived as a solution to unemployment, leading to permissive regulation \cite{anwartheconversation}, and such contexts are difficult to investigate \cite{glasius2018research}. Additionally, systematic concealment of human labour in AI, framed as business secrecy, perpetuates technological exceptionalism and prevents disclosure of working conditions \cite{newlands2021lifting, perrigo2023exclusive}.

Despite these obstacles, content moderators in Africa have organised collectively. However, effectiveness has been limited: short term contracts and the fact that BPOs do not face the same reputational pressures as technology companies constrain workers' impact \cite{muldoon2024typology}. Administrative and political obstacles hinder ``freedom of association, collective bargaining, and social dialogue,'' the first ILO category for measuring decent work in supply chains \cite{supplychainshub}. It is to document and counteract this censorship that the \emph{Data4Mods}'s project was established.
\section{Related Work}

Research on content moderators' working conditions has grown significantly, though the sector's secrecy makes this challenging. Whilst existing literature tends to treat AI data companies as exceptional entities, they operate as BPOs: a well-established organisational form analysable through traditional workplace studies. A crucial yet understudied element is the employment contract: to our knowledge, no existing research has systematically examined content moderators' contracts within BPOs, despite their centrality to the employment relationship. The following subsections outline the organisation of work in data work BPOs and what existing research tells us about working conditions, establishing the framework against which we compare the contracts recovered through workers' data access requests.

\subsection{BPO Work Organisation}

The organisational forms of AI data work are varied and depend on several criteria related to both the nature of the data work expected and the contracting parties. Based on observational fieldwork within a BPO specialising in AI data with sites in Kenya and Uganda, Muldoon and al.\cite{muldoon2024typology} identified different categories of AI data work institutions by examining the types of workers involved, i.e. self-employed, crowdsourced, or employees, and the types of services associated — and therefore data work — such as subcontractors, i.e. generalists or AI data-oriented, or technology companies that perform data work internally. Depending on the volume of data to be processed, confidentiality expectations, and specific knowledge and skills required (linguistic, technical, etc.), companies decide whether to process data internally or outsource it to entities such as BPOs.

More precisely, a BPO is defined as ``a form of outsourcing that involves contracting a third-party service provider to carry out specific parts of a company's operations'' \cite{miceli2022data}. While seemingly independent, BPOs are subject to the power of their clients, who control the duration of the partnership, specifications, and their execution \cite{miceli2022data}. The case of South African content moderator Daniel Motaung \cite{Motaung} illustrates this subordination: his demands for better working conditions triggered reprisals from Sama, its client Meta, and subcontractor Majorel, all acting in concert against workers who joined his protests.
The organisational model of AI data work BPOs mirrors BPOs in other sectors that practise the fragmentation of production work on an international scale \cite{ahmad2022moderating, tubaro2025does, anwar2025value}, with the same asymmetries and dependencies directly affecting working conditions.

\subsection{Working Conditions and Surveillance}

Surveillance and performance monitoring are central features of data work BPOs. Workers' time is strictly monitored: tasks must be completed within set time frames or achieve predefined accuracy rates, and failure to meet performance standards may result in pay cuts or dismissal. This monitoring also generates conflicts between workers regarding disagreements in data annotation \cite{abdelkadir2025role}, shaped by individual annotator characteristics \cite{diaz2018addressing, sang2022origin, sap2022annotators, pei2023annotator}. Hierarchical rigidity further compounds these conditions: professional figures such as quality analysts and subject matter experts are accorded more legitimacy than content moderators, even when the latter possess relevant knowledge \cite{abdelkadir2025role}. Abusive working conditions and the inability to influence platform policies are factors present in the BPO workplace that prevent data workers from fully expressing their expertise. These conditions contribute to well-documented occupational health problems, particularly mental health consequences from exposure to traumatic material \cite{gebrekidan2024content, workers2025we, cant2024feeding, steiger2021psychological, roberts2014behind, malgwi2025mental, arsht2018human, spence2023psychological, spence2024content, spence2025content, marsh2024overloaded}, compounded by insufficient breaks under intensive monitoring \cite{abdelkadir2025role}. Yet the contractual conditions that structure and perpetuate such health risks remain unexamined in the literature.

Beyond surveillance, precarity is enforced through legal instruments embedded in the employment relationship itself. Data work BPOs operate across both the Global South and the Global North \cite{miceli2024trains}: foreign workers in the latter are often subject to restrictive visa policies that tie their professional status to their employer, placing them in heightened subordination and discouraging them from speaking about their conditions. NDAs constitute another tool restricting employees' freedom of expression \cite{miceli2024trains}. As with their contested use during the MeToo movement to conceal misconduct \cite{grossi2025ndas}, NDAs in data work BPOs can function as a mechanism for client companies to discharge responsibility for their subcontractors' working conditions. It should be noted that confidentiality protocols may also include training and instructions that make work more manageable \cite{muldoon2024typology}, but this does not negate the broader concern that NDAs limit workers' ability to speak publicly about their conditions. Taken together, visa dependency and NDAs structurally suppress the transparency that employment contracts could otherwise provide.

\subsubsection{The Gap: Understanding Working Conditions through Contracts}

Despite growing documentation of data worker working conditions, existing research has relied primarily on interviews and surveys without systematically analysing employment contracts --- the key legal instrument that makes explicit and creates enforceable obligations regarding working hours, salaries, contract duration, and task definition. One notable exception is the Fairwork evaluation of working conditions at Sama, which included an analysis of employment contracts \cite{cant2023fairwork}. Assessed against five principles (Fair Pay, Fair Conditions, Fair Contracts, Fair Management, and Fair Representation), Sama initially failed to meet any of the thresholds. Following engagement with the Fairwork team, the company provided additional evidence and introduced 24 policy changes, achieving a score of 5/10 in 2023, which subsequently dropped to 3/10 in the 2024/2025 follow-up evaluation. Under Fair Contracts, the analysis revealed the prevalence of short rolling contracts, and Sama committed to making one-year contracts the default, or to aligning them with project timelines where possible. However, several limitations remain. As contract analysis was not the primary focus of the study, and evidence was obtained through voluntary company cooperation, the specific content of contractual clauses governing working conditions remains unexamined. Finally, the study did not extend to other major BPOs such as Teleperformance, included in our study. We extend this research by analysing employment contracts obtained through GDPR data access rights, a legally enforceable mechanism that compels disclosure of personal data held about workers, providing transparency and accountability into working conditions as a complementary view to the workers and companies' narratives.

\subsection{Personal Data Protection in the Workplace as a Research Tool}

Workplace surveillance has taken on a new dimension through its entanglement with AI and the collection of workers' personal data \cite{nguyen2021constant, abraha2025navigating, ball2010workplace}. As noted by the European Fundamental Rights Agency, the most advanced technologies for monitoring and controlling individuals' behaviour are mainly deployed in professional contexts \cite{european2010data}. Contemporary workplaces are characterised by sophisticated technologies tracking workers' actions and collecting extensive data on their performance, interactions, and completed tasks \cite{garden2018labor, custers2017worker} --- a dynamic particularly acute in data work BPOs, where monitoring is intensive and data collection extensive. Recent cases of data leakage illustrate that the risks this entails for workers are not merely theoretical \cite{scaleAIBI}: Scale AI, a US-based BPO whose clients include Google, Meta, OpenAI, Anthropic, Microsoft, and the US government, abruptly ceased operations in Kenya and Nigeria without warning its workers or paying them \cite{stahl2024kenyan}. The regulation of workplace data has grown considerably since the late 1970s, with data protection legislation now covering more than 160 countries \cite{greenleaf2023global}. Yet scholars have long questioned whether general data protection laws adequately address the distinct complexities of employment relationships \cite{simitis1999reconsidering, aloisi2025general}, concluding that omnibus regimes remain unsuited to the specific context of work.

Despite these limitations, the GDPR remains a ``pragmatic tool'' \cite{abraha2022pragmatic} that grants workers enforceable rights over their personal data. In Africa, workers remain largely unprotected under local frameworks, which are influenced by the GDPR and Convention 108+ but do not directly address workers' data \cite{abraha2025navigating, africaIAPP, dataprotectionafrica}. This is particularly significant given that data work is predominantly outsourced to BPOs in the Global South while the contracting companies operate in Europe \cite{williams2022exploited}. The GDPR's extraterritorial reach \cite{wolford2018does} --- an illustration of the ``Brussels effect'' \cite{bradford2020brussels} --- offers a practical remedy: where BPOs have a legal presence in the EU or process data flowing into or out of European territory, African workers can invoke European data protection law against their employers, as our methodology demonstrates.
\section{Methodology: Using EU GDPR to Audit Working Conditions in BPOs Outsourcing Labour in Africa} 

This study employed the \emph{Digipower methodology} \cite{digipower}: a data rights-based approach to investigate working conditions of data workers in a new setting: the ``AI supply chain'' \cite{muldoon2025politics}. This research was conducted as part of the project \emph{Data4Mods} developed by PersonalData.IO and the African Content Moderators Union (ACMU), in a context where content moderators are increasingly contesting their working conditions, pursuing legal cases against Meta for exploitative conditions via their subcontractors’ companies like Sama \cite{restofworldMetaModerators, siliconsavanna, guardianFacebookModerators, independentKenyanModerators}.

\subsection{Legal Framework: EU GDPR as a Labour Rights Instrument} 

The European General Data Protection Regulation (GDPR) is the primary legal framework governing the processing of personal data in the European Union. Under Article 4(1) GDPR, personal data means any information relating to an identified or identifiable natural person (i.e. the data subject). In an employment context, data access rights entitle workers, as data subjects, to request a copy of the data their employer holds about them. In practice, however, data controllers, i.e. the organisations responsible for processing personal data, do not apply the definition of personal data consistently, and the boundaries between personal data and employment data remain unclear, making variations in quantity and type of data returned common. This inconsistency is itself a finding of our research, as no worker has previously submitted a SAR to a data work BPO company.

The GDPR applies to any organisation established in the EU or processing the data of individuals located in the EU, regardless of where the organisation is headquartered. Less is known that if an organisation processes personal data flowing into or out of the EU, it is also subject to the GDPR regardless of where data subjects are located. This jurisdictional nexus is the cornerstone of our methodology: it enabled workers in African countries where data protection frameworks are weak or non-existent, to invoke, as data subjects, European data protection law against their employers. This cross-continental application of the GDPR represents both the novelty and the replicability of our approach.

The GDPR grants individuals a set of enforceable rights over their personal data, including the right to access, rectify, and port that data. The Digipower methodology \cite{digipower} draws upon two key provisions of the GDPR. Article 15 establishes the right of access, which entitles data subjects to obtain a copy of their personal data held by an organisation through a Subject Access Request (SAR). Under Article 12, organisations must respond within one month, with a possible two-month extension for complex cases. Importantly, online data portals provided by companies do not substitute for formal SARs, which allow data subjects to request specific data categories that automated tools may not capture. Article 20 establishes the right to data portability, which entitles data subjects to receive their personal data in a structured, commonly used, and machine-readable format, facilitating its transfer or personal use.

Previous research has theoretically examined how the GDPR can address labour rights and procedural problems on platforms, such as opaque ratings and automated decisions \cite{silberman2020using}. Few researchers have exercised these rights collectively \cite{mahieu2018}, and fewer still in the workplace, where existing studies focus solely on platform workers \cite{Perrig_2023, Pidoux_Dehaye_Gursky_2024, pidoux2025gaining, binns2025not}. In practice, ``interlegality'' \cite{li2022data}---the interaction of multiple legal frameworks such as data protection and labour rights---has already proven beneficial in litigation and labour rights claims. The legal grounding for treating employment data as personal data subject to GDPR rights is increasingly established in case law: in France, the Social Chamber of the French Supreme Court has clarified the place of the GDPR within the evidentiary framework of employment law disputes \cite{cass2025discrimination}, confirming that employment records constitute personal data over which workers hold enforceable rights. This is further evidenced by research documenting how Uber drivers in France and Switzerland recovered their personal data through SARs to contest algorithmic management \cite{Pidoux_Dehaye_Gursky_2024, pidoux2025gaining}. Our methodology applies this framework empirically and extends it in a significant direction. We focus on establishing transparency about the employment relationship itself in order to prove workers' rights. We do so in the AI supply chain, characterised by secrecy, where LLMs production depends on the labour conditions of content moderators, about whose contractual arrangements little is known. We employ data access and portability rights to recover foundational employment documents, such as labour contracts, and NDAs that were never provided to workers at the time of hiring, in order to analyse their working conditions through legal proofs.

\subsection{Co-Researcher Profiles and Sample Size Justification}

We supported five data workers (Table 1) in exercising their rights under the GDPR. These workers were based in Nigeria and Kenya and employed by Sama and Teleperformance. They participated as partners and co-researchers in the project rather than as research subjects, provided informed consent for the use of their data, and did not request anonymisation as they have already spoken publicly about their working conditions \cite{timeKaunaMalgwi}, \cite{timeMophatOkinyi}, \cite{timeJamesOyange}, \cite{kgomo2025}, \cite{timeRichardMathenge}.

\begin{table}[h]
\small
\caption{Co-Researcher Profiles}
\label{tab:profiles}
\begin{tabular}{p{0.5cm}p{1.8cm}p{2.8cm}p{6cm}}
\hline
\textbf{ID} & \textbf{Country} & \textbf{Employer} & \textbf{Position as per contract} \\
\hline
W1 & Kenya & Teleperformance & Swahili customer service representative grade A$^{**}$ \\[6pt]
W2 & Kenya & Teleperformance & Content Moderator$^*$ \\[6pt]
W3 & Kenya & Teleperformance & Customer service representative grade 1A$^{**}$ \\[6pt]
W4 & Nigeria & Sama & Content Moderator \\[6pt]
W5 & Nigeria & Sama & Content Moderator \\
\hline
\multicolumn{4}{l}{$^*$ Self-reported role; no contract was received in response to the SAR.} \\
\multicolumn{4}{l}{$^{**}$ Grade definition not provided in the contract.} \\
\end{tabular}
\end{table}

Three workers in Kenya submitted SARs to Teleperformance, whilst two workers in Nigeria submitted SARs to Sama. Teleperformance, formerly known as Majorel, is a multinational BPO company headquartered in Paris with offices across Europe and globally. The company provides customer service, content moderation, and other digital services to large technology platforms, including TikTok and Meta. Sama is a data annotation and AI training company headquartered in San Francisco with offices in the Netherlands. The company provides services to major technology firms, including Meta, Microsoft, Sony, Walmart, Getty Images, eBay, NASA, and Siemens, among others. Its workforce is engaged in tasks such as content moderation and data labelling. 

Four workers (W1, W3, W4, W5) received their contract and two (W4, W5) additionally received their NDA; our analysis is based on these documents respectively. The absence of a contract for W2, despite other workers at the same company receiving one, constitutes a GDPR violation and a research finding in itself.

This study does not claim statistical representativeness; it constitutes a proof-of-concept demonstrating that GDPR-based audits of employment conditions are legally viable and empirically useful for research on the AI supply chain. The five-worker sample is inherently limited by structural conditions of the field, including worker precarity and fear of retaliation, yet spanning two companies and two countries, and the evidentiary value of the recovered documents, represent a substantial achievement given workers' constraints.
\\
\subsection{Co-researching with Workers: SAR in Practice}

Submitting SARs by workers and analysing them implied six steps in practice.

\begin{enumerate}
    \item Workers searched on their former's employer's website information about their data rights, including privacy policies, applicable jurisdictions, the identity of the data protection officer, and procedures for submitting data requests. The means available for submitting SARs varied between the two companies and presented significant obstacles. Sama provided direct contact with their Data Protection Officer (DPO), enabling data subjects to submit requests via email correspondence. Teleperformance, by contrast, constrained them to use an online form that presented multiple barriers to the effective exercise of rights. The form offered only predefined options, making it impossible to simultaneously request access and data portability, thereby forcing data subjects to repeat the process. Furthermore, the form was connected to a generic, no-reply email address, preventing follow-up correspondence when responses were incomplete or when requests were closed without adequate justification. In one instance, Teleperformance answered by providing a new procedure to submit a SAR that included a broken URL, leaving the data subject unable to access the designated form and forcing them to navigate back to the same online form used before, creating misleading loops.
    According to the European Data Protection Board (EDPB) guidelines on deceptive design patterns \cite{deceptive2022}, such obstacles constitute infringements upon data subject rights: the broken link exemplifies a ``dead end,'' whilst the cumbersome repetition of requests exemplifies a ``longer than necessary'' pattern. These design choices create barriers that discourage the effective exercise of data protection rights under the GDPR; and the co-research process helped with not giving up and contact the DPO with legal means to contest such barriers. Outside the research project, data subjects without legal support would be unlikely to overcome them.
    
    \item {After identifiying the available SAR channels, workers documented the SAR process step by step and produced \href{https://wiki.personaldata.io/wiki/Main_Page}{tutorials} to share with other workers. 
    
    \item Prior to submitting requests, we had sessions where each worker documented the data they knew was being collected about them in the course of their daily work. 
    
    \item SARs were then drafted with the support of PersonalData.IO's Data Protection Officer, the template is \href{https://wiki.personaldata.io/wiki/Main_Page}{available} for other workers. SARs were personalised to each worker based on their documentation of data collected according to their experiences, and on data identified in each company's privacy policy. SARs were sent directly by each worker to the company's designated DPO by email or through the online form. Legally speaking, we adopted a strategy of graduated escalation. In the first instance, we requested that each company confirm whether it held personal data concerning the data subject. Once confirmed, workers submitted a more extensive request specifying the data categories of relevance. 
    
    \item Upon receiving responses, when personal data was denied or withheld, we submitted up to three follow-up requests, with a final request explicitly invoking workers' fundamental rights. This escalation only yield additional information for one worker (W4) at Sama. The process extended from November 2024 until March 2025, due to the companies' failure to provide complete data in a timely manner. Neither company furnished all requested information at the first stage, necessitating continuous follow-up correspondence comprising numerous emails and replies. This protracted engagement itself constitutes evidence regarding the practicality of exercising data access and portability rights for these companies' workers} in the context of cross-continental supply chains.
    
    \item Finally, we analysed qualitatively the data returned to identify information relevant to working conditions. Data analysis was conducted through individual sessions between a sociologist and each worker separately, as a data protection measure: workers decided not to all their personal data with one another and to analyse first their files individually. Workers selected which documents they wished to analyse and share. The sociologist then anonymised the selected files and constructed structured analytical tables in which key clauses and excerpts were extracted and compared across workers at the same company. The resulting tables were discussed collectively with workers to contrast the documentary evidence with their own experiences. 
    Our approach for assessing SAR responses is empirical and comparative. With three workers at Teleperformance and two at Sama, we compare what data was returned within and across companies. The variation itself --- a large volume of data returned for some workers, missing information for another one, or very little for others at the same company --- allows us to assess whether a response is complete or insufficient. Incompleteness is further analysed in light of workers' own experience of data they knew existed but did not find in the files returned to them.
\end{enumerate}

\section{Results}

\subsection{Personal Data obtained using EU GDPR in Africa}

Despite identical SAR requests submitted within each company, responses varied significantly both between and within companies. Teleperformance returned minimal data to W1, W2, and W3, whilst Sama provided considerably more detailed files to W4 and W5.

Teleperformance initially responded to W1 with only a Global Retention Policy rather than personal data. Following a second request submitted, all three workers received a link to access payslips. However, as none of them were still employed by the company, two had lost access to their login credentials with no mechanism provided to recover them, and the third managed to log in but found the payslips were not downloadable. Overall, personal data received were limited: W1 and W2 received a Global Retention Policy and a Data Collection Form, with W1 additionally receiving an Appointment Letter detailing employment conditions; W3 received a Data Collection Form, a Contract of Employment, and an Offer Letter. No formal contract was provided to W2.

It is important to highlight that, despite the incompleteness of the data transmitted, Teleperformance explicitly confirmed in an email responding to W1 that they retain employment data \textit{``in accordance with the Employment Act of Kenya, which requires that employment data is retained for five years post-employment termination, we have your employment data in our records.''} Furthermore, Teleperformance acknowledged in their response by email to W2 holding significantly more data than what was actually provided but that this was \textit{``deleted once an employee is deactivated from the system for a period of more than 12 or 20 months''} [including] \textit{``performance appraisals or reviews, training records and certifications completed, attendance records (clock-ins, absences), leave records (annual leave, sick leave, maternity/paternity leave), CCTV footage of me at workplace West-End offices, access logs (e.g. building entry logs or system access logs), IT usage records (e.g. browsing history and system usage).''}

This shows the extent of the data collected about workers that can be requested in future research for active workers and the utility of comparing requests across workers.

Sama provided more detailed personal data compared to Teleperformance, including a full list of personal data documents organised into four folders: \textit{1) Personal Information, 2) Employment Information, 3) Contracts, and 4) Background Checks}. A key finding is that all personal data returned by workers is directly relevant as employment data, and for the Sama case this covers information from the initial job application and recruitment process through to ongoing work performance. However, inconsistencies were observed despite both workers having similar profiles and roles within the company. W4 received significantly more personal information and background check documents, whilst W5 received more employment and training documents. Furthermore, in W4's case, some entries were deleted from the files returned: a performance rating, one page of the NDA, and payslips for the entire working period were not provided. Key files' discrepancies between W4 and W5 are detailed in Table \ref{tab:sama-comparison} in the Annex. While discrepancies may exist given the personal history of every worker in the company, the fact that fundamental documents are missing for some workers and not others within the same company point to a lack of respect of worker's rights.

\subsection{Analysis of Employment Contracts}

Among all the personal data received through the SAR process, our analysis focuses on three employment contracts and one appointment letter of four workers to understanding and evidencing content moderation working conditions. We examine more specifically six characteristics at Sama and Teleperformance: position titles, working hours, contract duration, task definition, salary transparency, and location flexibility.

\subsubsection{Position Titles Mismatching Actual Work}
The position titles recorded in the contracts show a significant discrepancy between the contractual designation and the actual work performed. At Teleperformance, workers were designated as \textit{``Swahili customer representative grade A''} (W1) or \textit{``Customer service representative grade 1A''} (W3), which definitions and scopes were not provided. In practice, W3 worked as a call centre agent for platform companies such as Zalando, additionally performing tasks as a content moderator. W1 worked as a content moderator and also as a ``feel-good manager'' training new workers, involving exposure to potentially harmful and traumatic material, despite having been told orally at the time of hiring that the role would only involve translating text from their local language. This mismatch between the stated position and actual tasks obscures the true nature of the work and may have implications for workers' access to appropriate occupational health support and legal protections specific to content moderation.

At Sama, the position title \textit{``Content Moderator''} appeared only in the offer letter and extension contract, not in the primary employment contract, and the work scope was not explained accurately as with W1. This inconsistent documentation means that the foundational contractual document governing the employment relationship failed to accurately specify the role for which the worker was hired.

\subsubsection{Unpredicatable Working Hours}

Both companies require workers to operate within a 24/7 environment with limited control over their schedules. Teleperformance contracts stipulate that employees work between 45 and 48 hours per week across shifts assigned by supervisors, with the explicit provision that \textit{``additional hours may be required based on business requirements''} and that \textit{``employees working on real time basis will have their working time regulated by their supervisors.''} This formulation grants management unilateral authority to modify working hours without employee consent. Sama's contracts contain similarly open-ended provisions, requiring employees to work \textit{``whichever shift he or she is assigned''} and, notably, to \textit{``devote the whole of their time and attention during business hours and at such other times as may be necessary''} without additional remuneration. This language effectively extends the employer's claim over workers' time beyond the contracted 45 hours, creating structural unpredictability.

\subsubsection{Precarious Contracts of Short-Term Duration}
The contracts reveal a pattern of short-term and renewable arrangements that maintain workers in a state of uncertainty. At Teleperformance, W3 received a three-month contract, whilst W1 received a one-year fixed-term contract explicitly described as one \textit{``which may or may not be renewed based on business needs/performance.''} Probationary period is extensive for W1: six months, and one month for W3, during which termination requires only 15 days' notice. At Sama, contracts were initially for one year but were subsequently renewed for progressively shorter periods—W3 had their contract renewed twice for one year, then for only one week, followed by a three-month extension. This pattern of diminishing contract duration exemplifies the precarity embedded in these employment relationships.

\subsubsection{Vague or Absent Task Definition}

The contracts are notably vague regarding the actual tasks workers are expected to perform. Teleperformance contracts reference \textit{``roles/duties provided for in the attached job description,''} yet the job description was not included in the data provided through the SAR. Furthermore, the contract explicitly states that \textit{``functions and duties may be altered at the discretion of management,''} granting the employer the ability to modify job requirements unilaterally. Sama's contracts fail to mention tasks altogether. Only W5's contract includes a document describing the \textit{``scope of work''}, yet it does so in vague and non-specific terms. Notably, the document never uses the words ``content moderation.'' Tasks are framed entirely in customer service language: workers are expected to \textit{``assist our community,''} \textit{``resolve inquiries empathetically,''} and \textit{``advocate for our community,''} whilst references to content moderation are concealed in phrases such as \textit{``reports of potentially abusive content''} and \textit{``monitoring reports of abuse on the site.''} There is no mention of the type of content workers will be exposed to, no reference to traumatic material, and no indication of the psychological risks involved, that should be disclosed before a worker accepts a position. At the same time, the document explicitly requires \textit{``high affinity and cultural awareness of political and social situations''} for the relevant market, confirming that the role demands specialised contextual and linguistic expertise. This gap between the competencies required and their absence from the job title, salary, and contractual protections illustrates the systematic undervaluation of data work. Moreover, this absence of clear task definition leaves workers without contractual protection against arbitrary changes to their responsibilities.

\subsubsection{Salary with Conditional Bonuses}

At Sama, the salary of 40,000 Ksh plus a 20,000 Ksh allowance appeared only in the offer letter (W4), not in the formal contract. W5 had the same salary and this information was in the contract. At Teleperformance, W1's salary was 30,000 Ksh and 36,000 Ksh for W3 (approx. \$230 USD), described for both as \textit{``inclusive of house allowance.''} Both companies tie a significant portion of potential earnings to performance bonuses—up to 20 per cent of monthly gross salary—conditional upon \textit{``achievement of set targets.''} However, these targets are not specified in the contracts, rendering a substantial portion of workers' expected income dependent upon opaque and potentially arbitrary performance metrics. It is worth noting that W5 had her annual salary review document specifying a \textit{performance rating of 0} but indicating at the same time an increase of \textit{1,400 Ksh}, leaving the worker with more uncertainties about how performance was measured.

\subsubsection{Location Flexibility Favouring the Employer}

Both companies reserve extensive rights to relocate workers. Teleperformance contracts state that workers \textit{``may be called upon to work at such other places as may be designated by management from time to time.''} Sama's contracts are even more expansive, requiring employees to \textit{``travel to such other places whether in or outside Kenya''} and explicitly reserving the company's right \textit{``to change the location of the Employee's place of work as may be reasonably necessary to respond to changing business needs.''} Their contracts assert that \textit{``the Employee's employment status and the other terms and conditions of his or her employment will not be adversely affected by any such requirements''}, thereby pre-emptively disclaiming any obligation to compensate workers for the disruption caused by relocation.

\subsubsection{NDAs}

Sama workers received NDAs as part of their employment documents, though these were not provided at the time of hiring. The NDAs are particularly restrictive and disproportionate for an individual employment context, creating a marked power asymmetry despite being framed as a \textit{``mutual''} non-disclosure agreement.
Three provisions are especially relevant to working conditions. First, the definition of confidential information extends far beyond standard business secrets to encompass employee information, personally identifiable information, and any information related to the company's operations (clause 1). In practice, this prevents workers from disclosing facts about their own employment situation, including working hours, tasks, and performance standards that are relevant to discuss for unionisation efforts. Second, confidentiality obligations survive contract termination for five years (clause 6), silencing workers long after they leave the company and limiting their ability to testify and organise collectively. Third, the injunctive relief provision (clause 7) entitles the company to seek court orders against workers, not merely financial compensation. For a precarious worker without legal support, the threat of injunctive proceedings constitutes a powerful deterrent against speaking out. Taken together, these provisions function less as mutual confidentiality protections than as instruments of labour control, structurally suppressing transparency about working conditions in a sector already characterised by opacity.
It is worth noting that in the initial data request, W4 received an incomplete version of the NDA with page 2 missing, meaning clauses 4 to 11 out of a total of 17 clauses were absent from the document; the complete version was only obtained following a second request. Second, whilst the two Sama workers were employed at the same location, their NDAs were issued by different legal entities: W4's contract names \textit{``Samasource Kenya Limited''} whilst W5's names \textit{``Samasource Kenya Private Limited.''} This discrepancy raises questions about corporate accountability and continuity of legal obligations when the contracting entity changes name.

\subsubsection{Contractual Provisions and their Impact on Working Conditions}

Taken together, these companies' contractual provisions reveal a systematic pattern of opacity and unpredictability that favours employer flexibility at the expense of worker security. Working hours are subject to unilateral modification; contract durations are short and uncertain; tasks are vaguely defined and subject to managerial discretion; salary components are conditional and incompletely documented; and location requirements grant employers extensive relocation rights. The absence of complete contractual documentation further compounds these issues, leaving workers without recourse to the basic terms of their own employment.

Workers' accounts confirm that these contractual conditions directly shaped their daily experience. Performance was continuously monitored, yet workers never received their evaluation scores nor any explanation of how metrics were calculated, creating a climate of opacity that generated constant anxiety. The pressure to process high volumes of content labelling tickets led workers to struggle between speed and accuracy, as mistakes in label selection subsequently affected their own performance records. When understaffing increased ticket volumes, workers were required to remain until all tasks were completed, sometimes until late at night when public transport had ceased, making it difficult and unsafe to return home. Workers also reported arriving at their workplace only to be reassigned to a different office location without prior notice, a practice consistent with the broad relocation clauses identified in the contracts.

Short-term contracts structurally suppressed workers' ability to assert their rights: as one co-researcher noted, the precarity of contract renewal meant that raising concerns carried no weight and risked termination. Workers were frequently assigned to train new colleagues, which they experienced as a signal that their positions could easily be filled, further discouraging them from speaking out. Finally, salary provisions proved as opaque in practice as on paper: some workers never received bonuses despite meeting performance targets, others received minimal amounts without explanation, and promotions were perceived as dependent on personal relationships with managers rather than on merit or documented performance, directly contradicting the contractual provisions analysed above.

\section{Discussion}
\subsection{What Employment Contracts Tell Us About Data Workers' Working Conditions}

Regarding position titles, workers hired as ``content moderation'' or ``customer service representative'' but performing content moderation and other tasks, without clear work scope definition, are denied recognition of specific occupational health risks associated with exposure to traumatic material, potentially rendering them ineligible for protections such as mandatory breaks or trauma informed counselling.

Regarding working hours, contracts stipulate 45 to 48 hour weeks but grant employers unilateral authority to require additional hours ``based on business requirements'' without additional compensation, creating ``permanent availability'' where workers cannot plan their lives. While research documents intensive time pressure within strict accuracy parameters \cite{abdelkadir2025role}, our analysis shows this monitoring operates within a broader temporal regime of employer control. These provisions contradict the flexibility narrative: workers are not granted flexibility, employers are. Regarding contract duration, short term rolling contracts maintain workers in perpetual insecurity that inhibits collective organising, including probationary periods of up to six months that reinforce vulnerability. The Sama worker whose contract was renewed for one week exemplifies arrangements designed to signal disposability. The Fairwork evaluation documented Sama's stated commitment to one year contracts \cite{cant2023fairwork} which was the case of W4, W5 in our study, yet other elements in the contract make this duration problematic, e.g. uncertainty about renewals, employer's full flexibility over the employment conditions.

Regarding task definition, salary, and mobility, contracts either omit task definitions or assert that duties ``may be altered at management's discretion'', which has particular significance where exposure to different harmful content has distinct psychological effects \cite{steiger2021psychological, roberts2014behind, spence2024content}. Salaries appear inconsistently, with portions dependent on unspecified targets. Relocation clauses enable companies to shift operations while workers bear disruptions. Moreover, NDAs further compound these structural disadvantages. Rather than protecting legitimate business secrets, these instruments function primarily to suppress transparency about working conditions and inhibit collective organising, reinforcing the same opacity that characterises the contracts themselves.

Finally, workers that did not receive their contract, NDA, or only an appointment letter, represent a denial of fundamental labour rights at BPOs. The employment contract represents a major development in workers' rights, breaking with forced labour norms. While it objectifies subordination, it can be seen as ``a structure of governance with democratic deficits'' \cite{davidov2025theory}, yet it enshrines worker freedom \cite{didry2013approche}, representing a ``symbolic revolution'' \cite{didry2021contrat}. As Bourdieu noted, ``with permanent employment and a regular salary, an open and rational awareness of time can be formed'' \cite{bourdieu1977}, cited in \cite{didry2021contrat}. Without contracts, workers cannot prove promised terms, challenge violations, or pursue remedies. Companies' acknowledgement through GDPR responses that contracts and NDAs exist, combined with their failure to provide these to workers, suggests strategic practices maintaining power asymmetry. This occurs within a specific policy context: Kenya has chosen to export workers abroad through agencies accused of exploitation \cite{semaforkenya}, while encouraging BPO sector growth \cite{kenyagov}, creating a policy environment favouring AI companies over worker protections. 

Our findings challenge the exceptionalism narrative that characterise discourse around the AI industry. The AI industry claims exceptionalism \cite{doorn2020price, vallas2020platforms} to justify departures from established regulatory frameworks, yet the opacity surrounding data work stems not from technical complexity but from deliberate organisational choices: there is no technical reason why workers cannot receive employment contracts or why tasks and salary structures cannot be transparently specified. 

Our GDPR-based methodology and contractual analysis document what companies actually do, as recorded in their own formal terms.

Research on building transparent and accountable AI systems \cite{miceli2021documenting, hutchinson2021towards} can be extended to labour practices within the AI supply chain. Just as academics have developed methods for auditing algorithmic bias and fairness \cite{raji2019actionable, raji2020closing, wieringa2020account}, equally rigorous methods can audit working conditions in AI production. Companies developing ``explainable AI'' could be held to equivalent standards regarding employment: performance metrics as transparent to workers as algorithmic decisions are claimed to be for end users. The asymmetry whereby companies demand trust in their systems while maintaining opacity regarding labour practices exposes the selective nature of industry commitments to transparency.

\subsection{Data Access Rights as an Accountability Method for Advocacy and Researching the AI Industry}

Our findings demonstrate that GDPR SARs constitute a legally enforceable method for establishing accountability in the AI industry, particularly within the opaque context of outsourced data work. Content moderators exercised their data access and portability rights themselves, compelling BPOs to disclose employment contracts and NDAs that had been systematically withheld, in some cases despite workers having signed them. This approach represents a methodological innovation: while existing research has relied on interviews, company statements, and observation \cite{roberts2014behind, steiger2021psychological, miceli2024trains, fairwork2025}, these methods cannot compel disclosure of formal legal documents. GDPR access thus offers researchers a legally grounded tool to pierce corporate opacity.
Through GDPR SARs, we obtained several categories of personal data that illuminate workplace data collection practices: payslips documenting remuneration, performance ratings revealing evaluation mechanisms, NDAs, employment contracts establishing precarity within formal terms, and records of data processors indicating surveillance practices (that we do not cover in this article). This demonstrates that employment related data qualifies as personal data under GDPR and can be retrieved through access requests, even when workers were not provided some documents upon hiring.

The utility of this method operates at three levels. First, it shifts the burden of documentation from workers to companies. GDPR requests compel companies to provide records that carry legal weight and can serve as evidence in labour disputes or regulatory proceedings. Second, this method identifies systemic patterns of precarity embedded in organisational structures. The inconsistencies we documented, including variations in contract duration, salary, and document completeness across workers in identical roles, expose structural problems that would be difficult to establish through interviews alone. The contracts, payslips, and performance records obtained through the SAR process constitute legal evidence that workers can mobilise in legal proceedings claiming labour rights. However, the restrictive NDAs identified in this study mean that access to appropriate legal support will be essential for workers seeking to use this evidence without exposing themselves to the risk of breach of contract claims. Companies acknowledged holding extensive data while failing to provide complete documentation, creating a record of non compliance that can support workers' claims, filing data protection complaints and inforing regulatory oversight. Third, this method operationalises the GDPR's extraterritorial reach where national labour protections are weak. Both Teleperformance and Sama initially refused disclosure, asserting that Kenyan law governed rather than the GDPR. Following our insistence based on the companies' European operations, both provided at least partial responses, demonstrating that invoking the GDPR triggered compliance and created enforceable rights for workers in the Global South whose labour supports European operations.

However, our findings reveal limitations. Companies employed strategies to obstruct disclosure: providing broken URLs, requiring access through platforms to which former employees lacked credentials, and initially asserting GDPR inapplicability. These are all expressions of companies' ability to circumvent the SAR framework by engaging in ``null compliance'' \cite{wright2024null}. Moreover, companies' selective definitions of ``personal data'' created barriers, with substantial variability in data received and strategic invocation of business confidentiality. This confirms that concerns raised regarding digital omnibus legislation \cite{noyb2024gdpr, datarights} reflect already existing practices whereby companies treat the definition of personal data as subject to their discretion, underscoring the need for clear, non restrictive definitions in employment contexts. 

Despite these limitations, data access rights remain valuable. Future research could build on this approach with a larger sample including currently employed workers, and conduct a legal analysis of working conditions through the lens of Kenyan labour law. The methodology is replicable wherever workers are employed by companies subject to the GDPR, and its potential to strengthen labour rights in cross-continental AI supply chains extends well beyond the present case. The combination of GDPR requests with union organising creates stronger leverage than traditional research methods alone: workers who collectively exercise their data rights are better positioned to challenge obstructive practices, and researchers who partner with civic organisations ensure that data access serves workers' interests rather than purely academic ones for the sake of data collection. Finally, computer scientists developing data infrastructure can contribute by designing systems that facilitate GDPR compliance from the outset, for instance by building SAR response functionalities, and ensuring worker-related data, falling under personal data, is stored in structured and retrievable formats.
\section{Conclusion}

Content moderators in the Global South face precarious working conditions characterised by opacity, precarity, and systematic denial of fundamental labour rights, yet remain largely undervalued within an industry that promotes narratives of technological exceptionalism. Workers who received incomplete contractual information have no means to prove promised terms or pursue legal remedies, especially if they are bound by restrictive NDAs that structurally prevent them from speaking about their conditions or organising collectively.
Our research demonstrates that GDPR data access rights constitute a legally enforceable method for auditing working conditions in the AI supply chain, uncovering the legal architecture of precarity embedded in employment relationships that companies would otherwise control entirely. This attention to foundational employment relationships extends AI ethics concerns beyond algorithmic fairness to encompass the human labour underpinning these systems, advocating for a future where workers possess the documentation necessary to assert their rights.
Crucially, our methodology positions workers as co-researchers exercising their own rights, ensuring that data access serves their interests and strengthens their capacity for collective action. Just as coal miners transformed their initially invisible and precarious conditions into the foundation for labour law and democratic institutions, content moderators organising collectively today may similarly shape the regulatory frameworks governing AI labour for generations to come.


\bibliographystyle{ACM-Reference-Format}
\bibliography{biblio}
\section{Endmatter}

\subsection{Acknowledgements}
We are deeply grateful to Nawale Lamrini from PersonalData.IO, whose legal expertise in data protection enabled the realisation of this work. We also thank the reviewers for their time and valuable feedback.

\subsection{Generative AI usage statement}

We used generative AI (Claude) to improve readability, condense text and correct grammar, as the authors are non-native English speakers. No original research text, conceptual frameworks or empirical analyses were generated by AI.

\section{Appendix}

\begin{table}[h]
\small
\caption{Comparison of SAR Data Received by W4 and W5 (Sama)}
\label{tab:sama-comparison}
\begin{minipage}{0.48\textwidth}
\setlength{\tabcolsep}{4pt}
\begin{tabular}{p{3cm}p{1cm}p{1cm}}
\hline
\textbf{Document} & \textbf{W4} & \textbf{W5} \\
\hline
\multicolumn{3}{l}{\textit{Personal Information}} \\
Payslips (some) & \checkmark & \checkmark \\
E-Permit / Work Permit & \checkmark & \checkmark \\
Alien ID & \checkmark & \checkmark \\
Travel Passport & \checkmark & \checkmark \\
Curriculum Vitae & \checkmark & \checkmark \\
KRA Pin Certificate & $\times$ & \checkmark \\
Police Clearance Certificate & \checkmark & $\times$ \\
Passport photo & \checkmark & $\times$ \\
Security Bond & \checkmark & $\times$ \\
Birth Certificate & \checkmark & $\times$ \\
Academic Certificates & \checkmark & $\times$ \\
Beneficiary Nomination Form & \checkmark & $\times$ \\
Candidate Information Form & \checkmark & $\times$ \\
Employee Emergency Data Form & \checkmark & $\times$ \\
Dependants Capture Form & \checkmark & $\times$ \\
Onboarding Forms & $\times$ & \checkmark \\
\hline
\multicolumn{3}{l}{\textit{Background Checks}} \\
Education and Employment Check & \checkmark & $\times$ \\
Criminal Check & \checkmark & $\times$ \\
Certificate of Verification & \checkmark & $\times$ \\
Check Report & $\times$ & \checkmark \\
\hline
\end{tabular}
\end{minipage}
\hfill
\begin{minipage}{0.48\textwidth}
\setlength{\tabcolsep}{4pt}
\begin{tabular}{p{3cm}p{1cm}p{1cm}}
\hline
\textbf{Document} & \textbf{W4} & \textbf{W5} \\
\hline
\multicolumn{3}{l}{\textit{Employment Information}} \\
Annual Salary Review Letter & \checkmark & \checkmark \\
Warning Letter & \checkmark & \checkmark \\
SamaHome Program Agreement & \checkmark & \checkmark \\
Work from Home Rules \& Guidelines & \checkmark & \checkmark \\
Confirmation Letter & \checkmark & \checkmark \\
Homebase Declaration Form & \checkmark & \checkmark \\
Activity Code Refresher Training Form & \checkmark & \checkmark \\
PIP Acknowledgement Form & \checkmark & \checkmark \\
Redundancy Notice & \checkmark & \checkmark \\
Redundancy Exit Letter & \checkmark & \checkmark \\
GPL8 Induction Training Programme & $\times$ & \checkmark \\
SamaU Certificates GPL8 & $\times$ & \checkmark \\
Work from the Office Procedures & $\times$ & \checkmark \\
Expatriate \& Company Responsibility Guidelines & $\times$ & \checkmark \\
\hline
\multicolumn{3}{l}{\textit{Contracts}} \\
Contract of Employment & \checkmark & \checkmark \\
GPL8 Contract Extension & \checkmark & \checkmark \\
NDA (Mutual Non-Disclosure Agreement) & \checkmark & \checkmark \\
Signed Offer Letter & \checkmark & $\times$ \\
Letter of Offer & \checkmark & $\times$ \\
Expatriate \& Company Responsibility Guidelines & \checkmark & $\times$ \\
GPL8 Contract Extension Template & $\times$ & \checkmark \\
\hline
\multicolumn{3}{l}{\checkmark = received \quad $\times$ = not received} \\
\end{tabular}
\end{minipage}
\end{table}

\end{document}